\documentclass[12pt]{article}

\usepackage[english]{babel}
\usepackage[utf8x]{inputenc}
\usepackage{amsmath}
\usepackage{graphicx}
\usepackage[colorinlistoftodos]{todonotes}
\usepackage{hyperref} 
\usepackage{color}
\usepackage{amssymb}
\usepackage[caption=false]{subfig}
\usepackage[margin=1in]{geometry}
\hypersetup{
    colorlinks,
    citecolor=blue,
    filecolor=blue,
    linkcolor=blue,
    urlcolor=blue
}
\usepackage{multirow}

\usepackage{natbib}
\setcitestyle{authoryear,aysep={,}}

\newcommand\floor[1]{\lfloor#1\rfloor}
\newcommand\ceil[1]{\lceil#1\rceil}

\begin{document}
\title{Mixed Effect Dirichlet-Tree Multinomial for Longitudinal Microbiome Data and Weight Prediction} 
\author{Yunfan Tang and Dan L. Nicolae \\
{\normalsize\emph{University of Chicago }}} 
\date{}
\maketitle
\begin{abstract}
Quantifying the relation between gut microbiome and body weight can provide insights into personalized strategies for improving digestive health. In this paper, we present an algorithm that predicts weight fluctuations using gut microbiome in a healthy cohort of newborns from a previously published dataset. Microbial data has been known to present unique statistical challenges that defy most conventional models. We propose a mixed effect Dirichlet-tree multinomial (DTM) model to untangle these difficulties as well as incorporate covariate information and account for species relatedness. The DTM setup allows one to easily invoke empirical Bayes shrinkage on each node for enhanced inference of microbial proportions. Using these estimates, we subsequently apply random forest for weight prediction and obtain a microbiome-inferred weight metric. Our result demonstrates that microbiome-inferred weight is significantly associated with weight changes in the future and its non-trivial effect size makes it a viable candidate to forecast weight progression.
\end{abstract}

\section{Introduction}
Next-generation technologies in DNA sequencing have vastly expanded our understanding of microbiome and how it impacts the health condition of human host. Since the initial endeavors from Human Microbiome Project \citep{turnbaugh2007}, researchers have been able to associate microbial compositions with a number of diseases such as inflammatory bowel diseases \citep{kostic2014} and type-2 diabetes \citep{hartstra2015}, as well as identify particular taxa as biomarkers for these phenotypes. As an integral component of immune system and metabolic activities, microbiome is regarded as a promising candidate for personalized medicine \citep{elrakaiby2014,shukla2015}.

Technological advances in sequencing contrast with much slower development in statistical analysis methods. Data output from 16s ribosomal RNA sequencing pipeline typically present major statistical challenges such as compositional data, variability in sequencing depth, overdispersion,  relations among taxa and localized signals \citep{li2015,thorsen2016}. Few statistical algorithms are tailored to all of these new characteristics. Directly applying existing methods such as support vector machine and random forest \citep{pasolli2016} can lead to loss of prediction accuracy or results hard to interpret. Recent focus on tackling these difficulties involves decomposing the overall community data according to the structure of their phylogenetic tree \citep{tang2016,silverman2017,wang2017}. These transformations untangle the high-dimensional compositional nature of microbiome data so that conventional analytical tools can be directly applied. A particularly interesting class of models is Dirichlet-tree multinomial (DTM) \citep{dennis1991}, which extends the traditional Dirichlet multinomial (DM) onto phylogenetic trees and provides greater flexibility. DTM naturally incorporates sequencing depth, overdispersion and can be easily adapted to deal with localized signals. Application of DTM to microbial analysis has been shown to yield noticeable improvements for detecting phenotype-microbiome associations \citep{tang2016} and in prediction accuracy \citep{wang2017}. 

Gut microbiome has been linked to body weight/BMI in a number of human and animal studies \citep{sweeney2013,lecomte2015}, although the exact mechanism is yet to unfold. Microbiome is known to be highly sensitive to diet \citep{david2014}, but diet alone does not always lead to weight change in the absence of certain species \citep{fei2013,thaiss2016}. These studies have so far disproportionately focused on obesity traits and largely neglected how microbial variability interacts with weight fluctuations for healthy subjects. Motivated by the innovative microbiome-predicted age metric in a recent study on Bangladesh newborns \citep{subramanian2014}, we seek to provide insights on microbiome-weight relationship by defining microbiome-inferred weight on a healthy cohort comprised of newborns up to 2 years old. Our algorithm removes the unwanted effects from covariates based on a mixed effect DTM model and employs multi-scale empirical Bayes shrinkage for improved estimation of microbial proportions, both of which are designed to cater to the unique characteristics of microbiome data. We use random forest to predict weight using these shrinkage estimators as explanatory variables. Microbiome-inferred weight encodes the microbial information into an interpretable summary that is capable of forecasting future weight trajectories.

The rest of the paper is organized as follows. Section \ref{sec:DTM regression on microbiome} contains a brief review of DTM setup followed by elaboration on mixed effect DTM. It then presents empirical Bayes shrinkage and simulation results. Section \ref{sec:Weight prediction for Bangladesh newborns} builds on the shrinkage residuals to predict newborns' weight in the Bangladesh dataset and demonstrate that microbiome-inferred weight is capable of forecasting short-term weight fluctuations. Section \ref{sec:Discussion} concludes this paper with possible future work.

\section{DTM regression on microbiome data} \label{sec:DTM regression on microbiome}
\begin{figure}[h]
\centering\includegraphics[height=7cm]{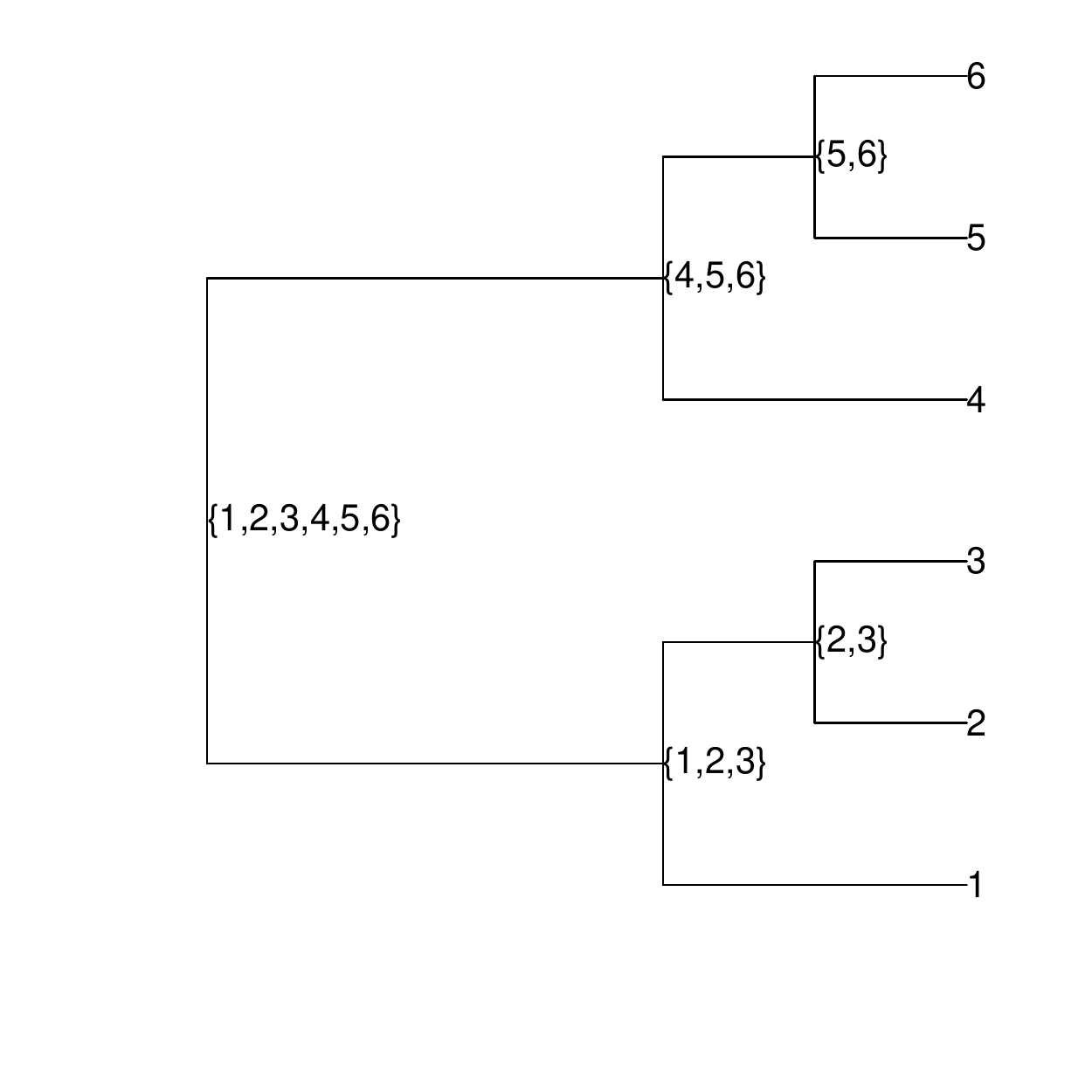}
\caption{\label{fig:Phylotree example} An example of a phylogenetic tree with six OTUs. Each internal node is uniquely labeled with the set of OTUs underneath it.}
\end{figure}

Here we briefly review the DTM framework as in \cite{tang2016}. Let $\mathcal{T} = (\Omega, \mathcal{I})$ be a rooted phylogenetic tree where the set of operational taxonomic units (OTU) $\Omega$ are placed on the leaves and $\mathcal{I}$ is the set of all internal nodes. Without loss of generality, we assume $\Omega = \{1, 2, ..., K\}$ where $K = |\Omega|$. We represent the elements in $\mathcal{I}$ to be subsets of $\Omega$ since each internal node is uniquely identified by the subset of all OTUs underneath it. Figure \ref{fig:Phylotree example} shows an example of a simple phylogenetic tree with 6 OTUs and 5 internal nodes. This tree has $\Omega = \{1,2,3,4,5,6\}$ and $\mathcal{I} = \big\{ \{1,2,3,4,5,6\}, $ $\{1,2,3\}, \{4,5,6\}, \{2,3\}, \{5,6\} \big\}$. Also $\forall A \in \mathcal{I}$, define $c(A) \in \mathcal{I}$ and $d(A) \in \mathcal{I}$ to be the first and second child node of $A$, respectively. In the example above, we can define $c(\{1,2,3\}) = \{1\}$ and $d(\{1,2,3\}) = \{2,3\}$. The ordering of first and second child under each internal node is completely arbitrary and does not affect the tree structure. By definition, $A = c(A) \cup d(A)$ and $c(A) \cap d(A) = \emptyset$.

Now consider the longitudinal microbial dataset of Bangladesh newborns \citep{subramanian2014}. Backgrounds of this dataset are described in Section \ref{sec:Weight prediction for Bangladesh newborns}. Let $x_{ij} = (x_{1,ij}, x_{2,ij}, ..., x_{K,ij})$ be the $j$th microbial observation in $i$th family, where $1\leq i \leq m$ and $1 \leq j \leq n_i$ with $m$ being the total number of families and $n_i$ being the number of observations in $i$th family. Notice that $x_{ij}$ can come from any child in $i$th family. Every $x_{ij}$ is a $K$-dimensional count vector representing the number of sequences in each of the $K$ OTUs. For each internal node $A \in \mathcal{I}$, the total count of $A$'s descendant OTUs in $x_{ij}$ is $x_{A,ij} = \sum_{\omega \in A} x_{\omega,ij}$, since $A$ is represented as a subset consisting of all its descendant OTUs. In particular, $x_{\Omega,ij}$ is the total number of sequences (sequencing depth) in observation $(i,j)$. The Dirichlet-tree multinomial (DTM) model has the following hierarchical representation for all $A$: 
\begin{equation*}\label{DTM}
q_{A,ij} \overset{\text{i.i.d.}}{\sim} \text{Dir}\big(\nu_A \psi_A, \nu_A(1-\psi_A)\big), \hspace{3mm} x_{c(A),ij} | x_{A,ij}, q_{A,ij} \sim \text{Binomial}(x_{A,ij}, q_{A,ij})
\end{equation*}
where $\psi_A \in (0,1)$ is the mean proportion of the counts in $c(A)$ over counts in A, and $\nu_A > 0$ is a dispersion parameter that governs the level of variation across samples. Without incurring any confusion, $q_{A,ij}$ denotes the value on the first dimension of the outcome from a two-dimensional Dirichlet distribution (since elements from both dimension sum up to 1). All of the Dirichlet priors and the conditional binomial distributions are mutually independent. We only explicitly model counts on $A$'s first child since by definition, $x_{c(A),ij} + x_{d(A),ij} = x_{A,ij}$. 

\cite{dennis1991} showed that DTM degenerates to the global DM distribution when the following condition is satisfied for all $A \in \mathcal{I}$: $\nu_A \psi_A = \nu_{c(A)}$ if $c(A) \in \mathcal{I}$, and $\nu_A(1-\psi_A) = \nu_{d(A)}$ if $d(A) \in \mathcal{I}$. This means that DM is nested in the DTM family. \cite{tang2016}, using a likelihood ratio test, compared DTM and DM in the American Gut dataset \citep{mcdonald2015}  and concluded DTM provides significantly improved fit over global DM. 

\subsection{Mixed effect DTM} \label{sec:Mixed effect DTM}
We next model the association of microbial proportions with covariates through a logit link. Both age and sex effects are assumed to be fixed, and the family effect is assumed to be random. Let $t_{ij}$ be the child's age at time of observation and $s_{ij}$ be the indicator variable for sex. For each $A \in \mathcal{I}$, we assume
\begin{equation}  \label{eq:M1}
u_{A,i} \overset{\text{i.i.d.}}{\sim} N(0, \sigma_A^2) 
\end{equation}
\begin{equation}  \label{eq:M2}
\log \frac{\psi_{A,ij}}{1-\psi_{A,ij}} | u_{A,i} = \beta_{A,0} + \beta_{A,1} t_{ij} + \beta_{A,2} s_{ij} + u_{A,i}
\end{equation}
\begin{equation} \label{eq:M3}
q_{A,ij} | \psi_{A,ij} \sim \text{Dir}\big(\nu_A \psi_{A,ij}, \nu_A (1-\psi_{A,ij}) \big) 
\end{equation}
\begin{equation} \label{eq:M4}
x_{c(A),ij} | x_{A,ij}, q_{A,ij} \sim \text{Binomial}(x_{A,ij}, q_{A,ij})
\end{equation}
where $u_{A,i}$ is the $i$th random family effect on $A$, $\beta_{A,0}$ is the intercept, $\beta_{A,1}$ is the age effect and $\beta_{A,2}$ is the sex effect. Let $\beta_A = (\beta_{A,0}, \beta_{A,1}, \beta_{A,2})$ be the vector of intercept and fixed effects for short, which is shared across all samples on node $A$.

The family effect $u_{A,i}$ contains all the unknown factors that would alter the gut microbiome, such as shared environment (diet, hygiene) and genetics. As in (\ref{eq:M1}), the family effect is assumed to follow normal distribution with a different standard deviation $\sigma_A$ on each node $A$. The distributions in (\ref{eq:M1})-(\ref{eq:M4}) are mutually independent both within and across internal nodes. This longitudinal DTM breaks down the global distribution of all taxa counts into independent local components, each modeled through its own set of parameters. 

Let $\theta_A = (\beta_A, \nu_A, \sigma_A)$ be the parameters associated with $A$. Define 
$$x_{A} = (x_{A,11}, x_{A,12}, ..., x_{A,1n_1}, x_{A,21}, ..., x_{A_2n_2}, ..., x_{A, mn_m} )$$ 
as the vector of all observations on $A$. Similarly, define $u_A = (u_{A,1}, u_{A,2}, ..., u_{A,m})$ be the vector of all family random effects. The conditional density of $x_{c(A)}$ is therefore
\begin{align*} \label{eq:den1}
f_{\theta_A}(x_{c(A)} | x_A) &= \int f_{\theta_A}(x_{c(A)} | x_A, u_A) \phi_{\sigma_A}(u_A) d u_A \nonumber \\
&= \prod_{i=1}^{m} \int_{-\infty}^{\infty} \phi_{\sigma_A}(u_{A,i}) \prod_{j=1}^{n_i} f_{\theta_A}(x_{c(A),ij} | x_{A,ij}, u_{A,i}) d u_{A,i}, \text{ by independence of $u_{A,i}$'s}
\end{align*}
where
\begin{equation} \label{eq:den2}
f_{\theta_A}(x_{c(A),ij} | x_{A,ij}, u_{A,i}) = {{x_{A,ij}}\choose{x_{c(A),ij}}} \frac{ \big(\nu_A\psi_{A,ij}\big)^{\uparrow x_{c(A),ij}} \big(\nu_A(1-\psi_{A,ij})\big)^{\uparrow x_{d(A),ij}} }{\nu_A^{\uparrow x_{A,ij}}}
\end{equation}
is the DM density under the notation $\alpha^{\uparrow k} =  \prod_{l=0}^{k-1} (\alpha+l)$, and $\phi_{\sigma_A}(\cdot)$ is the normal density with mean 0 and standard deviation $\sigma_A$. The log likelihood of $\theta_A$ is
\begin{align} 
l(\theta_A) &= \log f_{\theta_A}(x_{c(A)} | x_A) \nonumber \\
&= \sum_{i=1}^{m} \Big( -\frac{1}{2} \log \sigma^2_A + \log \int_{-\infty}^{\infty} \exp \Big\{ -\frac{u_{A,i}^2}{2\sigma_A^2 } + \sum_{j=1}^{n_i} \log f_{\theta_A}(x_{c(A),ij} | x_{A,ij}, u_{A,i}) \Big\} d u_{A,i}  \Big) \nonumber \\ 
&= \sum_{i=1}^{m} \Big( -\frac{1}{2} \log \sigma^2_A + \log \int_{-\infty}^{\infty} \exp \Big\{ -\frac{u_{A,i}^2}{2\sigma_A^2 } + \sum_{j=1}^{n_i} l_{ij}(\theta_A) \Big\} d u_{A,i}  \Big) \label{eq:loglik}
\end{align}
up to an irrelevant constant, where $l_{ij}(\theta_A) = \log f_{\theta_A}(x_{c(A),ij} | x_{A,ij}, u_{A,i})$. Since distributions on different internal nodes are independent, optimization of $\theta$ can be executed separately. For each node $A$, we use gradient based optimization to obtain MLE $\hat\theta_A = \text{argmax}_{\theta_A} l(\theta_A)$. See Appendix for details of optimization.

\subsection{Removing covariate effects and empirical Bayes shrinkage}
Predicting weight from microbiome for newborns present major statistical challenges since microbial compositions evolves with age for newborns \citep{subramanian2014} and can be related to sex. Microbiome is also associated with a number of latent factors such as diet \citep{tang2016}, genetics \citep{goodrich2014}, hygiene, etc. Subjects from different family can demonstrate distinct microbial profile due to these latent factors yet have similar weight. In order to optimize prediction performance, it is crucial to remove these effects from microbiome data. Our model (\ref{eq:M1})-(\ref{eq:M4}) presents a framework to account for the effect of age ($\beta_{A,1}$), sex ($\beta_{A,2}$) and family ($u_{A,i}$). As mentioned before, the family random family effect is interpreted as the sum of all contributions from diet, genetics, etc. Under this DTM framework, we can use the estimated coefficients and predicted random family effect to remove these extraneous effects. 

In addition to impacts from the aforementioned covariates, the inferred microbial proportion is also prone to variabilities of sequencing depth. To clarify, there have been two types of proportions used for microbial analysis: OTU proportions $x_{\omega,ij}/x_{\Omega,ij}$ and internal node proportions $x_{l(A),ij} / x_{A,ij}$, the latter calculated from a phylogenetic representation. For OTU proportions, subsampling has been a popular technique to offset variations in $x_{\Omega,ij}$ but is clearly sub-optimal as it discards useful information \citep{mcmurdie2014}. For internal node proportions, variability of $x_{A,ij}$ among the samples is even greater than $x_{\Omega,ij}$ since it is not only affected by sequencing depth but also individual microbial compositions whenever $A \neq \Omega$. Typically, the node count $x_{A,ij}$ can vary in several order of magnitude ranging from zero (i.e. complete missing data) to hundreds of thousands. Our goal is to incorporate variability of $x_{A,ij}$ into a valid statistical estimation procedure that does not throw away any useful data. This is achieved through a multi-scale empirical Bayes shrinkage on the observed node proportions towards the estimated mean. Empirical Bayes shrinkage has been widely used for signal processing \citep{johnstone2005,xing2016}. One of its most desirable properties is its ability to adjust the degrees of shrinkage based on data, which avoids the issue of prior specification. When the data admits any type of hierarchical decomposition such as wavelet transformation, empirical Bayes can be applied to each layer of distribution separately, thus achieving multi-scale shrinkage. We naturally extend this idea to DTM framework by individually shrinking each local DM distribution. After obtaining the shrinkage estimate of node proportions, we then subtract effects of age, sex and family to obtain residuals. 

For a fixed internal node $A$, suppose $(i,j)$ is the observation to be shrunk. The procedure for estimating $q_{A,ij}$ and calculating residual $r_{A,ij}$ involves empirical best prediction \citep{jiang2001} of random family effect based on data collected prior to $t_{ij}$. This rolling algorithm proceeds as follows:

\begin{enumerate}
\item For each $A$, calculate MLE $\hat{\theta}_A$ by maximizing (\ref{eq:loglik}) using either a separate training dataset, or the current dataset but only with observations collected prior to sample $(i,j)$.
\item If $x_{A,ij'} = 0$ for all $j'\leq j$ (i.e. no prior data in $i$th family), set $\hat u_{A,i} = 0$. Otherwise, predict the family random effect $u_{A,i}$ by empirical best prediction, using all available data in $i$th family collected before $t_{ij}$ in the current dataset: 
\begin{align}
\hat{u}_{A,i} &= E_{\hat{\theta}_A} \big(u_{A,i} | \{ (x_{c(A),ij'}, x_{A,ij'})| j' \leq j \} \big) \nonumber \\
&= \frac{\int u_{A,i} \prod_{j' \leq j} f_{\hat{\theta}_A}(x_{c(A),ij'} | x_{A,ij}, u_{A,i}) \phi_{\hat\sigma_A}(u_{A,i}) d u_{A,i} }{\int \prod_{j' \leq j} f_{\hat{\theta}_A}(x_{c(A),ij'} | x_{A,ij}, u_{A,i}) \phi_{\hat\sigma_A}(u_{A,i}) d u_{A,i}} \label{eq:EP}
\end{align}
\item Calculate the empirical Bayes estimate of $q_{A,ij}$:
\begin{equation}  \label{eq:EBshrink1}
E_{\hat{\theta}_A}(q_{A,ij} | \hat u_{A,i}, x_{c(A),ij}, x_{A,ij}) = \frac{x_{c(A),ij} + \hat\nu_A \hat\psi_{A,ij}}{x_{A,ij} + \hat\nu_A}, 
\end{equation}
where
$$\hat\psi_{A,ij} = \frac{e^{\hat\beta_{A,0} + \hat\beta_{A,1}t_{ij} + \hat\beta_{A,2}s_{ij} + \hat u_{A,i}}}{1+ e^{\hat\beta_{A,0} + \hat\beta_{A,1}t_{ij} + \hat\beta_{A,2}s_{ij} + \hat u_{A,i}}}$$
\item Remove the effect of age, sex and family to obtain the residual
\begin{align} 
r_{A,ij} &= E_{\hat{\theta}_A}(q_{A,ij} | \hat u_{A,i}, x_{c(A),ij}, x_{A,ij}) - E_{\hat{\theta}_A}(q_{A,ij} | \hat u_{A,i}) \nonumber \\
&= \frac{x_{c(A),ij} + \hat\nu_A \hat\psi_{A,ij}}{x_{A,ij} + \hat\nu_A}   - \hat\psi_{A,ij} \label{eq:EBshrink2}
\end{align}

\end{enumerate}
Integrations in (\ref{eq:EP}) are calculated using the same set of techniques mentioned in the Appendix. The posterior estimate in (\ref{eq:EBshrink1}) shrinks the observed node proportion $x_{c(A),ij}/x_{A,ij}$ towards the estimated mean $\hat \psi_{A,ij}$ depending on total node count $x_{A,ij}$ and dispersion levels. In the case of complete missing data (i.e. $x_{A,ij} = 0$), this yields $E_{\hat{\theta}_A}(q_{A,ij} | \hat u_{A,i}, x_{c(A),ij}, x_{A,ij}) = E_{\hat{\theta}_A}(q_{A,ij} | \hat u_{A,i}) = \hat \psi_{A,ij}$ and thus $r_{A,ij} = 0$. In other words, empirical Bayes shrinkage is a natural extension of imputing missing data by the mean for DM distribution. 

\subsection{Simulation}
Here we demonstrate the effectiveness of empirical best prediction in (\ref{eq:EP}) and empirical Bayes shrinkage estimator (\ref{eq:EBshrink1}) through a simple simulation. Since our algorithm operates on each node separately, it suffices to simulate data on just one node $A$. We set $\beta_A = (-1, 0.1, 0.2), \nu_A = 10$ and $\sigma_A = 0.5$ as the true parameter values (assume $t_{ij}$ is measured in hundreds of days). We generate data for $m = 10$ families, all of them having twins with different sex. Each child has his/her longitudinal microbial observations recorded on a grid of 15 time points equally spaced from 0.1 (10 days) to 8 (800 days). For each observation, the total node count $x_{A,ij}$ is first drawn from a negative binomial distribution with mean 100 and dispersion 0.2 (so its variance is $100 + 100^2/0.2 = 50100$), followed by mixed effect DTM (\ref{eq:M1})-(\ref{eq:M4}) to generate $x_{c(A),ij}$. We choose the parameters of the negative binomial distribution as described in order to mimic the extreme variability of $x_{A,ij}$ in real microbiome dataset.

We simulate the entire longitudinal dataset for a total of 1000 runs. In each run, the empirical best prediction (\ref{eq:EP}) and empirical Bayes shrinkage (\ref{eq:EBshrink1}) are invoked at the third, sixth, ninth, twelfth and the fifteenth observation time point, respectively. For example, the third observation time point is 1.2285, i.e. 122.85 days. All observations taken equal or prior to this time point are used to produce MLE $\hat\theta_A$ and subsequently calculate $\hat u_{A,i}$ and $\hat q_{A,ij} = E_{\hat{\theta}_A}(q_{A,ij} | \hat u_{A,i}, x_{c(A),ij}, x_{A,ij})$ for all $(i,j)$. The prediction mean square error (MSE) for random family effect $u$, defined as $\sum_{i=1}^{10} (u_{A,i} - \hat u_{A,i})^2/10 $, is used to benchmark empirical best prediction. For empirical Bayes shrinkage, we compare its MSE against the binomial proportion $\tilde q_{A,ij} = x_{c(A),ij} / x_{A,ij}$ and use the difference of their MSE $ \sum_{i,j} \big( (\tilde q_{A,ij} - q_{A,ij})^2 - (\hat q_{A,ij} - q_{A,ij})^2 \big)$ as performance metric. About 30\% of simulated $x_{A,ij}$ are equal to zero and hence $\tilde q_{A,ij}$ is unobtainable. We ignore the square error in those cases. 

\begin{figure}[h!]
\centering
\subfloat{\includegraphics[height=7cm]{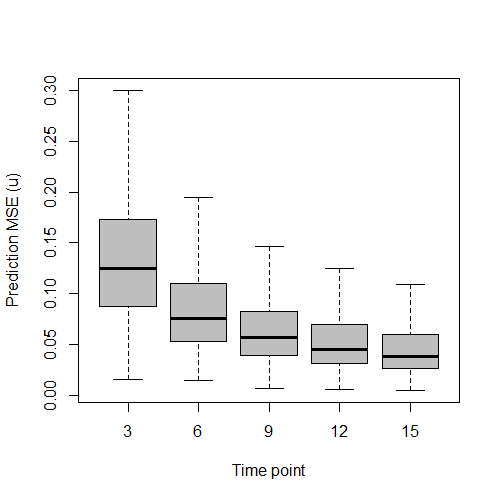}} 
\subfloat{\includegraphics[height=7cm]{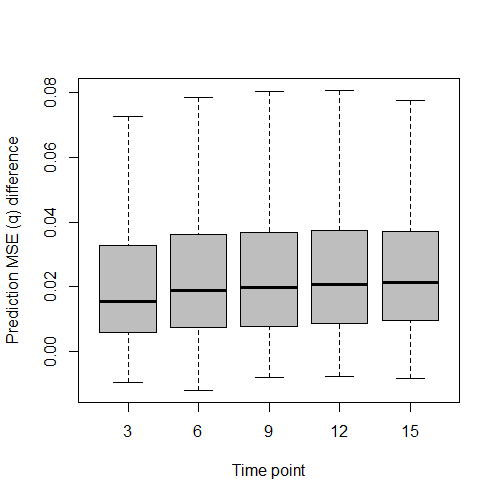}}
\caption{\label{fig:u_boxplot} Box plot of prediction MSE for random family effect (left) and difference of prediction MSE between empirical Bayes estimator and binomial proportion (right). Estimations are invoked at the third, sixth, ninth, twelfth and fifteenth time point from a total of 1000 runs.}
\end{figure}

Figure \ref{fig:u_boxplot} presents the box plots of MSE calculated at each of these aforementioned time points from the simulated data, showing improved prediction performance for both estimators as time increases. Notice that ignoring observations with $x_{A,ij} = 0$ is equivalent to assigning the value of $\hat q_{A,ij}$ to $\tilde q_{A,ij}$ in those cases, so MSE improvement of $\hat q_{A,ij}$ is arguably most influenced by observations with small but non-zero $x_{A,ij}$.

\section{Weight forecast for Bangladesh newborns} \label{sec:Weight prediction for Bangladesh newborns}

Although the past decade has witnessed burgeoning efforts devoted to microbial research, longitudinal study of the interaction between microbiome and body weight has been mostly lacking. Here, we use the dataset from a nutritional study of Bangladesh children from \cite{subramanian2014}. The study contains anthropometric measurements and fecal microbiome samples periodically for newborns up to two years old living in an urban slum in Dhaka, Bangladesh. Fecal microbiome samples (V4-16S rRNA) were sequenced on Illumina MiSeq platform, which generates $26,580\pm 26,312$ (mean $\pm$ s.d.) reads per sample. We obtained the assembled reads from the authors' website and further processed the sequencing data through QIIME v1.9.1 \citep{caporaso2010} under default settings. OTUs were picked by open-reference method with Greengenes reference database (version 13\_8).

\subsection{Microbiome-inferred weight}
The twin/triplet healthy cohort consists of longitudinal observations of newborns from 12 families. 11 of these families have twins and the other family has triplets. Each observation includes fecal microbiome sample and optional weight measurement from a certain newborn. Samples collected within 7 days of antibiotic administration are excluded from subsequent analysis. This yields a total of 382 microbial observations, 324 of which are accompanied with weight measurements. We select top 100 OTUs with highest counts, excluding those with more than 95\% of their counts occurred in a single observation. The final 100 OTUs selected make up more than 94.8\% of all sequence counts. For weight, we use the weight-for-age corrected z-score, abbreviated as WAZ \citep{who2009}, as the response. WAZ is obtained through subtracting raw weight by the mean value of the reference population at given age and sex followed by standardizing the residuals.

Next, we apply empirical Bayes shrinkage with leave-one-family-out cross validation (i.e. 12 fold) to calculate $r_{A,ij}$ in (\ref{eq:EBshrink2}). As the name suggests, each CV fold leaves out all samples from a certain family as test data and use the remaining samples as training data. The MLE estimate $\hat\theta = \{\theta_A: A \in \mathcal{I} \}$ is obtained from optimizing (\ref{eq:loglik}) separately for each $A$ on all training samples. After that, $\hat\theta$ is applied on test samples to calculate $r_{A,ij}$ in a rolling fashion according to (\ref{eq:EBshrink2}). 

\begin{figure}[h]
\centering
\subfloat{\includegraphics[height=7cm]{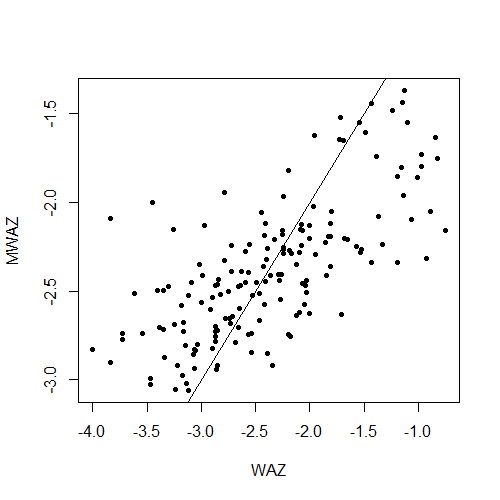}} 
\subfloat{\includegraphics[height=7cm]{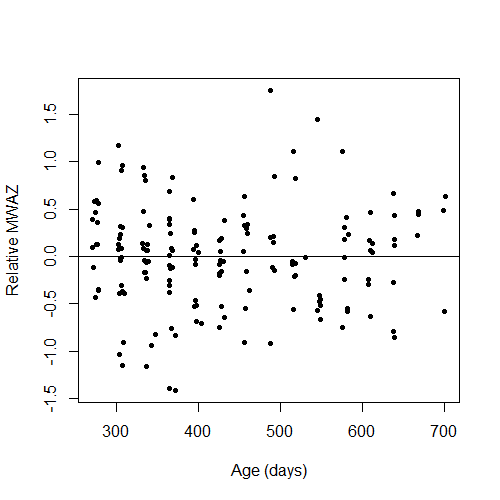}}
\caption{\label{fig:delta vs time} Left: scatter plot of MWAZ ($\hat y_{ij}$) vs observed WAZ ($y_{ij}$), with the solid line drawn at intercept 0 and slope 1. Right: scatter plot of relative MWAZ ($e_{ij} = \hat y_{ij} - y_{ij}$) vs age ($t_{ij}$).}
\end{figure}

Let $y_{ij}$ be the WAZ of observation $(i,j)$ and suppose elements in $\mathcal{I}$ are assigned an arbitrary order as $A_1, A_2, ..., A_{|\mathcal{I}|}$. We train the random forest with $(r_{A_1,ij}, r_{A_2,ij}, ..., r_{A_{|\mathcal{I}|},ij})$ as predictor and $y_{ij}$ as response for $1\leq i \leq m$ and $1 \leq j \leq n_i$, using the \verb ranger  function in R package \verb ranger  with $5 \times 10^4$ trees and $|\mathcal{I}|/3$ variables randomly sampled as candidates at each split. In order to have adequate amount of prior data to be used in (\ref{eq:EP}), only observations with $t_{ij} \geq 250$ are predicted. This yields a total of 165 training samples for random forest regression. Define $\hat y_{ij}$ as the out-of-bag prediction \citep{breiman2001} and $e_{ij} = \hat y_{ij} - y_{ij}$ as the residuals. Using out-of-bag prediction avoids the need for an independent validation dataset. We call $\hat y_{ij}$ microbiome-inferred WAZ (MWAZ) and $e_{ij}$ relative MWAZ. The random forest regression gives prediction MSE $\sum_{i,j} (y_{ij} - \hat y_{ij})^2/\sum_i n_i = 0.286$. As a comparison, using mean response $\bar y = \sum_{i,j} y_{ij}/\sum_i n_i $ as predictor yields MSE equal to 0.529. In other words, random forest predictor $\hat y_{ij}$ reduces MSE of mean predictor by $46.0 \%$. In Figure \ref{fig:delta vs time} we present the MWAZ vs WAZ plot and relative MWAZ vs age plot. With the exception of a few observations at around $t_{ij} = 500$, variability of $e_{ij}$ gradually decreases as $t_{ij}$ increases and it further stabilizes at $t_{ij} > 400$. This is consistent with the fact that estimates of $u_{A,i}$ become more accurate as we accumulate more prior data.

Through the \verb rfcv  function implemented in the \verb randomForest  package, we calculate that using top 10 internal nodes with highest importance yields the smallest prediction MSE. The importance of internal node $A$ is measured by increase in prediction MSE after permuting $r_{A,ij}$ in all out-of-bag samples, averaged over all trees. For each one of these top 10 nodes with highest importance, we provide taxonomies for both of its children in Table \ref{table:node taxa}. Since QIIME only outputs taxonomy assignments for leaf OTUs, we use a simple majority-vote rule to determine taxonomy for internal nodes. At any fixed rank, the taxonomy of a certain internal node is resolved if more than 80\% of the counts in its descendant OTUs have the same taxa on that rank. For example, both children of top node in Table \ref{table:node taxa} have more than 80\% of their counts belonging to phylum Firmicutes. Its first child has than 100\% of its counts in class Bacilli, but its second child only has 77.2\%. Therefore on class level, the algorithm classify the left child as Bacilli but the right child as unresolved (indicated by a dash). 

The reason we provide taxa for both children in Table \ref{table:node taxa} is that each local DTM distribution is conditioned on total node count $x_{A,ij}$, according to (\ref{eq:M4}). Therefore, changes in first and second child counts are restrained to be complimentary (if one decreases, the other always increases). Any discovered signal on the internal nodes should be attributed to relative level of counts on the  first vs second child, but neither one in particular. Notice that in some cases, the taxonomic resolution of both children can be drastically different, such as the top ranked node. This happens when one of the children is a single OTU but the other contains a wide variety of OTUs with distinct taxa. If there is enough prior evidence of the taxonomic level with which weight is mostly likely associated, then we can reasonably locate the signal to whichever child closer to that target level.

\begin{table}[h!] 
\scriptsize
\centering
\begin{tabular}{ c c c c c c}
\hline
Importance & Phylum & Class & Order & Family & Genus \\
\hline
\multirow{2}{*}{0.146} & \multirow{2}{*}{Firmicutes} & Bacilli & Lactobacillales & Lactobacillaceae & Lactobacillus \\
&  & --- & --- & --- & --- \\
\multirow{2}{*}{0.095} & \multirow{2}{*}{Actinobacteria} & \multirow{2}{*}{Actinobacteria} & \multirow{2}{*}{Bifidobacteriales} & \multirow{2}{*}{Bifidobacteriaceae} & \multirow{2}{*}{Bifidobacterium} \\
&  &  &  &  &  \\
\multirow{2}{*}{0.039} & \multirow{2}{*}{Firmicutes} & Bacilli & Lactobacillales & Lactobacillaceae & Lactobacillus \\
&  & --- & --- & --- & --- \\
\multirow{2}{*}{0.026} & \multirow{2}{*}{Firmicutes} & \multirow{2}{*}{Bacilli} & \multirow{2}{*}{Lactobacillales} & \multirow{2}{*}{Streptococcaceae} & \multirow{2}{*}{Streptococcus} \\
&  &  &  &  &  \\
\multirow{2}{*}{0.023} & \multirow{2}{*}{Firmicutes} & Erysipelotrichi & Erysipelotrichales & Erysipelotrichaceae & Eubacterium \\
&  & Clostridia & Clostridiales & --- & --- \\
\multirow{2}{*}{0.021} & \multirow{2}{*}{Actinobacteria} & \multirow{2}{*}{Coriobacteriia} & \multirow{2}{*}{Coriobacteriales} & \multirow{2}{*}{Coriobacteriaceae} & \multirow{2}{*}{---} \\
&  &  &  &  &  \\
\multirow{2}{*}{0.018} & \multirow{2}{*}{Firmicutes} & Bacilli & Lactobacillales & Leuconostocaceae & Leuconostoc \\
&  & --- & --- & --- & --- \\
\multirow{2}{*}{0.014} & \multirow{2}{*}{Firmicutes} & Clostridia & Clostridiales & Clostridiaceae & --- \\
&  & --- & --- & --- & --- \\
\multirow{2}{*}{0.009} & \multirow{2}{*}{Actinobacteria} & Actinobacteria & Bifidobacteriales & Bifidobacteriaceae & Bifidobacterium \\
&  & Coriobacteriia & Coriobacteriales & Coriobacteriaceae & --- \\
\multirow{2}{*}{0.009} & \multirow{2}{*}{Actinobacteria} & \multirow{2}{*}{Actinobacteria} & \multirow{2}{*}{Bifidobacteriales} & \multirow{2}{*}{Bifidobacteriaceae} & \multirow{2}{*}{Bifidobacterium} \\
&  &  &  &  & \\
\hline
\end{tabular}
\caption{\label{table:node taxa} Taxonomy classification for children of top 10 weight-predictive internal nodes in the order of decreasing importance from top to bottom row. Duplicated taxon of children nodes are condensed into a single entry. For example, both children of the top ranked node belongs to firmicutes phylum. Missing or unresolved taxa are indicated by dashes.}
\end{table}

\subsection{Using MWAZ for weight forecast}
Sensitivity of microbiome with respect to diet \citep{david2014} makes it a potential precursor to weight fluctuations. We demonstrate in this section that the relative MWAZ, $e_{ij}$, is capable of predicting weight changes in the future. If MWAZ is higher/lower than WAZ, then it is likely that the subject will exhibit increased/decreased weight in the near future. To verify this hypothesis, we first quantify the short-term change of WAZ per day $\delta_{ij}$ as follows:

$$\delta_{ij} = \frac{y_{ik} - y_{ij}}{t_{ik} - t_{ij}} \text{, where $k$ is the smallest index such that } 5 \leq t_{ik} - t_{ij} \leq 30,$$
where we use 30 days as a cutoff for short term weight changes. The 5 days minimum is to avoid $\delta_{ij}$ only capturing day-to-day random fluctuations. A total of 68 samples have future weights collected within $[5,30]$ days and thus their $\delta_{ij}$'s are obtainable. Within these 68 samples, $\delta_{ij}$'s have mean $2.20\times 10^{-3}$ and standard deviation 0.012, and $e_{ij}$'s have mean 0.073 and standard deviation 0.517.

\begin{figure}[h!]
\centering\includegraphics[height=7cm]{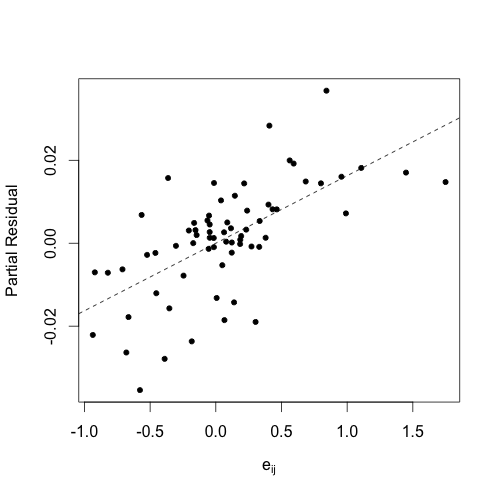}
\caption{\label{fig:PartialRes} Partial residual plot of $e_{ij}$ (relative MWAZ) in linear model (\ref{eq:pred delta}). Partial residual is calculated as $\delta_{ij} - \hat a_0 - \hat a_2 y_{ij} - \hat a_3 t_{ij} - \hat a_4 z_{ij}$. Dashed line is drawn at intercept 0 and slope $\hat a_1$.}
\end{figure}

In order to predict $\delta_{ij}$ from $e_{ij}$, we include a number of additional covariates  to reflect the best knowledge of weight development up to date. Our list of covariates includes current weight $y_{ij}$, current age $t_{ij}$ and backward per-day change of weight $z_{ij}$. $z_{ij}$ is defined similar to $\delta_{ij}$ except that the constraint $5\leq t_{ik} - t_{ij} \leq 30$ is replaced by $t_{ik} - t_{ij} < 0$.
We use a simple linear model to predict $\delta_{ij}$ as follows:
\begin{equation}  \label{eq:pred delta}
E(\delta_{ij}) = a_0 + a_1 e_{ij} + a_2 y_{ij} + a_3 t_{ij} + a_4 z_{ij}
\end{equation}
where $(a_0, a_1, ..., a_4)$ are the regression coefficients. This linear model yields $R^2 = 0.239$ and adjusted $R^2 = 0.191$. Estimate of the coefficient of $e_{ij}$ is $\hat{a}_1  = 0.016$ with standard error $4.51 \times 10^{-3}$ and p-value $6.00 \times 10^{-4}$. Of all other covariates, only $y_{ij}$ has its coefficient significant at 0.05 level. We also provide a partial residual plot of $e_{ij}$ in Figure \ref{fig:PartialRes}. To assess the contribution of $e_{ij}$ towards overall fit, we run another the model using the same set of covariates but without $e_{ij}$ in the predictor. This gives us $R^2 = 0.082$ and adjust $R^2 = 0.039$. Compare it with the result from (\ref{eq:pred delta}), including $e_{ij}$ alone yields improvement of unadjusted $R^2$ by $0.157$ and adjusted $R^2$ by $0.152$.

Next, we fit the model (\ref{eq:pred delta}) to subgroups by age and sex to inspect whether there is heterogeneity among the subgroups. For age, we use 400 days as the threshold since this appears to be where variability begins to stabilize in Figure \ref{fig:delta vs time}, except for the outliers around $t_{ij} = 500$. Sample sizes of these four subgroups are 33($t_{ij} \leq 400$), 35($t_{ij} > 400$), 20(male) and 48(female). The result is presented in Table \ref{table:subgroup fit}. We include $R^2$ from both with $e_{ij}$ and without $e_{ij}$ while keeping all other covariates $y_{ij}$, $t_{ij}$ and $z_{ij}$. Unsurprisingly, the highest increments of $R^2$ is obtained at $t_{ij} > 400$ group due to better stabilized $e_{ij}$'s. Results from male and female are also fairly consistent.

\begin{table}[h!] 
\small
\centering
\begin{tabular}{ c c c c c c}
\hline
Subgruop & $t_{ij} \leq 400$ & $t_{ij} > 400$ & Male & Female \\
\hline
$\hat a_1$ & 0.017 & 0.016 & 0.017 & 0.019\\
$R^2$ (w/o $e_{ij}$) & 0.077 & 0.130 & 0.107 & 0.132\\
$R^2$ (w/ $e_{ij}$) & 0.140 & 0.331  & 0.308 & 0.264  \\
\hline
\end{tabular}
\caption{\label{table:subgroup fit} Summary of linear model fit on different subgroups. Models with (third row) or without (second row) $e_{ij}$ have the same set of covariates $y_{ij}, t_{ij}$ and $z_{ij}$.}
\end{table}

\section{Discussion} \label{sec:Discussion}
We have demonstrated that random forest prediction of weight (MWAZ) using empirical Bayes shrinkage estimators based on mixed effect DTM model is linked with future weight progression. This sheds light on the interplay between microbial composition and body weight from a prediction perspective. While our procedure is carried out on a healthy cohort of newborns, we envision that application to malnourished population can be far more useful. In the same study of Bangladesh newborns \citep{subramanian2014}, the authors provided an additional cohort of children suffering from severe acute malnutrition. These children went through short-term therapeutic food interventions and had their weight measured before, during and after the food interventions. Although most subjects demonstrate noticeable weight increment during the treatment, there exists great variability in their weight progression after the treatment. Some individuals quickly lapsed into the same malnourished state, while some others have steadily increasing weight that could last for a few months. We suspect that gut microbiome plays a central role in determining how well the subject responds to food intervention. Forecasting weight response can provide treatment guidelines in terms of length and strength in order to restore subjects' weight to a proper level. Unfortunately, the wide usage of antibiotics on these treatment subjects makes it impossible to apply our MWAZ metric for weight prediction. Further studies need to be conducted in order to collect enough eligible samples for a thorough investigation of how microbiome impacts weight recovery.

\bibliography{ref}

\appendix 
\section*{Appendix: Optimization details}
Using (\ref{eq:den2}), $l_{ij}(\theta_A)$ is expressed as
\begin{equation}\label{eq:loglik2}
l_{ij}(\theta_A) = \sum_{\xi=0}^{x_{c(A),ij}-1} \log (\nu_A\psi_{A,ij} + \xi) + \sum_{\xi=0}^{x_{d(A),ij}-1} \log \big(\nu_A(1-\psi_{A,ij}) + \xi\big) - \sum_{\xi=0}^{x_{A,ij}-1} \log (\nu_A + \xi)
\end{equation}
up to an irrelevant constant, with the value of $\psi_{A,ij}$ given in (\ref{eq:M2}). Due to the summation operation, evaluating each $l_{ij}(\theta_A)$ involves $\mathcal O(x_{A,ij})$ computation cost. For high-throughput sequencing data, $x_{A,ij}$ could easily reach tens of thousands especially for $A$ close to the root. This can dramatically slow down numerical integration, which needs to evaluate the integrand on a dense grid. To overcome this difficulty, notice that all summations in (\ref{eq:loglik2}) take the form $g(\alpha,k) = \sum_{\xi=0}^{k-1}\log(\alpha+\xi)$. We use Taylor series approximation to fast compute $g(\alpha,k)$ for arbitrary values of $\alpha$ and $k$ based on expansion on integer grids.

To start, let $[\alpha]$ denote the closest integer to $\alpha$ and define $\epsilon = \alpha - [\alpha]$. Also, let $\floor{\cdot}$ and $\ceil{\cdot}$ be the floor and ceiling operators, respectively. When $[\alpha] + \xi > 0$, $\log(\alpha+\xi)$ can be expanded as 
\begin{equation}\label{eq:Taylor}
\log(\alpha+\xi) = \log([\alpha]+\xi) + \frac{\epsilon}{[\alpha]+\xi} - \frac{\epsilon^2}{2([\alpha]+\xi)^2} + ... 
\end{equation}

In order to simply the demonstration, suppose only quadratic expansion is used. We first calculate the values of the following three functions for all $k = 2, ..., x_{\Omega,ij}$ and store them into memory:
\begin{equation} \label{eq:series}
S_0(k) = \sum_{\xi=1}^{k-1} \log(\xi), \hspace{2mm} S_1(k) = \sum_{\xi=1}^{k-1} \frac{1}{\xi}, \hspace{2mm} S_2(k) = \sum_{\xi=1}^{k-1} \frac{1}{\xi^2}  
\end{equation}

Next, we choose an integer $T$ such that the approximation (\ref{eq:Taylor}) is invoked only when $\alpha + \xi \geq T$. This turns the original function value $g(\alpha,k)$ into 
\begin{align*}
g(\alpha,k) &= \sum_{\xi=0}^{\floor{T-\alpha}} \log(\alpha+\xi) + \sum_{\xi=\ceil{T-\alpha}}^{k-1} \log(\alpha+\xi) \nonumber \\
&\approx \sum_{\xi=0}^{\floor{T-\alpha}} \log(\alpha+\xi) + \big(S_0([\alpha] + k) - S_0([\alpha]+\ceil{T-\alpha}) \big) + \epsilon\big(S_1([\alpha] + k) \nonumber \\
&\hspace{5mm} - S_1([\alpha]+\ceil{T-\alpha}) \big) - \frac{\epsilon^2}{2}\big(S_2([\alpha] + k) - S_2([\alpha]+\ceil{T-\alpha}) \big) \label{eq:approx}
\end{align*}
assuming that $\alpha$ is not an integer. With $\mathcal O(\max_{i,j} x_{A,ij})$ memory complexity, calculating $g(\alpha,l)$ only has $\mathcal O(1)$ time complexity. As a result, (\ref{eq:loglik2}) is approximated by 
\begin{equation*} \label{eq:loglikapp}
l_{ij}(\theta_A) \approx g(\nu_A \psi_{A,ij}, x_{c(A),ij}) + g\big(\nu_A (1-\psi_{A,ij}), x_{d(A),ij}\big) - g(\nu_A, x_{A,ij})
\end{equation*}
In practice, we choose $T=10$ and fourth order Taylor expansion, which yields relative error less than $10^{-8}$, 

Another issue of calculating the log likelihood arises out of limitations in machine precision. The integrand in (\ref{eq:loglik}) includes product of $n_i$ likelihoods, each of which originally has a binomial coefficient as in (\ref{eq:den2}). Omission of the binomial coefficient does not affect the MLE estimate and avoids $\mathcal O (x_{A,ij})$ time complexity, but doing so yields extremely small values in the right side of (\ref{eq:loglik2}). With extremely small integrands, most numerical integration algorithms will not converge properly or report false estimates of integration errors. An easy fix to this problem is to introduce an additive constant on the exponent of (\ref{eq:loglik}) to bring its value back to normal ranges. We first choose an initial estimate $(\tilde{\beta}_A, \tilde\nu_A)$ and then use the following expression instead of (\ref{eq:loglik}) for log likelihood:
\begin{equation}
l(\theta_A) = \sum_{i=1}^{m} \Big( -\frac{1}{2} \log \sigma^2_A + \log \int_{-\infty}^{\infty} \exp \Big\{ -\frac{u_{A,i}^2}{2\sigma_A^2 } + \sum_{j=1}^{n_i} \big(l_{ij}(\theta_A) - l_{ij}(\tilde\theta_A)\big) \Big\} d u_{A,i}  \Big) \label{eq:loglik_const}
\end{equation}
where $\tilde\theta_A = (\tilde\beta_A, \tilde\nu_A, 0)$. The additive constant on the exponent will only increase the log likelihood by a constant and does not change its gradient. Due to large sequencing depth, it is advised to choose $(\tilde\beta_A, \tilde\nu_A)$ reasonably close their respective MLE, since otherwise $l_{ij}(\hat\theta_A) - l_{ij}(\tilde \theta_A)$ can be too large and jeopardize numerical precision of integration. In our implementation, $(\tilde{\beta}_A, \tilde\nu_A)$ is determined by MLE while fixing $\sigma_A$ = 0, i.e. no random effect present. Under such circumstance, the log likelihood simply comes from the DM distribution and thus optimization is trivial.

To calculate the gradient of (\ref{eq:loglik}) or equivalently (\ref{eq:loglik_const}) with respect to $\theta_A$, we simply need to switch the differential and integral operations. The partial derivative with respect to $\sigma_A$ is very straightforward and hence omitted. Other partial derivatives are as follows

\begin{equation} \label{eq:partial2}
\nabla_{(\beta_A,\nu_A)}l(\theta_A) = \sum_{i=1}^m \frac{\int_{-\infty}^{\infty} \exp \Big\{ -\frac{u_{A,i}^2}{2\sigma_A^2 } + \sum_{j=1}^{n_i} l_{ij}(\theta_A) \Big\} \sum_{j=1}^{n_i} \nabla_{(\beta_A,\nu_A)} l_{ij}(\theta_A) d u_{A,i}} {\int_{-\infty}^{\infty} \exp \Big\{ -\frac{u_{A,i}^2}{2\sigma_A^2 } + \sum_{j=1}^{n_i} l_{ij}(\theta_A) \Big\} d u_{A,i} }
\end{equation}
with
\begin{equation} \label{eq:partial3}
\nabla_{\nu_A} l_{ij}(\theta_A) = \psi_{A,ij} \sum_{\xi=0}^{x_{c(A),ij}-1} \frac{1}{\nu_A\psi_{A,ij} + \xi} + (1-\psi_{A,ij}) \sum_{\xi=0}^{x_{d(A),ij}-1} \frac{1}{\nu_A(1-\psi_{A,ij}) + \xi} - \sum_{\xi=0}^{x_{A,ij}-1} \frac{1}{\nu_A + \xi}  
\end{equation}
\begin{equation} \label{eq:partial4}
\nabla_{\beta_A} l_{ij}(\theta_A) = \frac{\nu_A C_{ij}}{(1+e^{\gamma_{A,ij}})(1+e^{-\gamma_{A,ij}})}\Big( \sum_{\xi=0}^{x_{c(A),ij}-1} \frac{1}{\nu_A\psi_{A,ij} + \xi} - \sum_{\xi=0}^{x_{d(A),ij}-1} \frac{1}{\nu_A(1-\psi_{A,ij}) + \xi} \Big)
\end{equation}
where $C_{ij} = (1, t_{ij}, s_{ij})$ and $\gamma_{A,ij} = \beta_A^T C_{ij} + u_{A,i}$. 

Both techniques for calculating the log likelihood introduced above are applicable to gradient calculation. First, we use the same approximation strategy with $\mathcal O(1)$ time complexity on the form $h(\alpha,k) = \sum_{\xi=0}^{k-1} 1/(\alpha + \xi)$ that are present in (\ref{eq:partial3}) and (\ref{eq:partial4}), using the following expansion on integer grids:
\begin{equation*}
\frac{1}{\alpha+\xi} = \frac{1}{[\alpha]+\xi} - \frac{\epsilon}{([\alpha]+\xi)^2} + \frac{\epsilon^2}{([\alpha]+\xi)^3} + ...
\end{equation*}
as long as $[\alpha] + \xi > 0$. Since approximation strategy of $h(\alpha,k)$ is highly akin to $g(\alpha,k)$, details are omitted. Second, we insert a same additive constant into the exponent on both the numerator and denominator of (\ref{eq:partial2}) for stable numerical behaviors.

Certain internal nodes exhibit constantly increasing log likelihood as $\sigma_A$ approaches to zero. Small values of $\sigma_A$ make normal density $\phi_{\sigma_A}(\cdot)$ close to Dirac function and can easily disrupt numerical integrations. As a countermeasure, we set a lower bound $\sigma_A \geq 10^{-3}$ on the optimization. When this lower bound is achieved, the objective function value (\ref{eq:loglik_const}) is compared with the log likelihood under the same $(\beta_A, \nu_A)$ but with $\sigma_A = 0$, i.e. a simple DM log likelihood. If the latter value is larger, we fix $\sigma_A = 0$ and proceed to optimize $(\beta_A, \nu_A)$.

We use the function \verb pcubature  implemented in R package \verb cubature  for numerical integration and the low-storage BFGS optimization \citep{liu1989limited} implemented in R package \verb nloptr  to calculate MLE.

\end{document}